\documentclass[12pt,epsf]{article}

 \usepackage{epsfig}
 \usepackage{times}
 \usepackage{epstopdf}
 \usepackage{amsmath}
 \usepackage{amssymb}
\usepackage{color}

\newcommand{\be}{\begin{equation}}
\newcommand{\ee}{\end{equation}}
\def\({\left (}
\def\){\right )}
\def\[{\left [}
\def\[{\right ]}

\begin{document}
\begin{titlepage}
\bigskip
\rightline
\bigskip\bigskip\bigskip\bigskip
\centerline {\Large \bf {Singularities in Hyperscaling Violating Spacetimes}}
\bigskip\bigskip
\bigskip\bigskip

\centerline{\large Keith Copsey$^{a,b}$ and Robert Mann$^a$}
\bigskip\bigskip
\centerline{${}^a$ \em Department of Physics and Astronomy, University of Waterloo, Waterloo, Ontario N2L 3G1, Canada}

\vspace{0.2 cm}

\centerline{${}^b$ \em Perimeter Institute for Theoretical Physics, Waterloo, Ontario N2L 2Y5, Canada}

\vspace{0.3cm}

\centerline{\em kcopsey@perimeterinstitute.ca, rbmann@sciborg.uwaterloo.ca}
\bigskip\bigskip

\begin{abstract}
We examine spacetimes which generalize Lifshitz scaling to allow hyperscaling violation invariance (i.e. a constant conformal transformation) for the types of singularities frequently found in the Lifshitz case.   We find that most of these spacetimes suffer either from a traditional naked curvature singularity or a null curvature singularity robust against stringy higher derivative corrections.  We do find a one-parameter family that evades these issues but this family requires exotic matter, in particular involving a negative energy density that may not be interpreted as due to a cosmological constant.

\end{abstract}
\end{titlepage}

\baselineskip=16pt
\setcounter{equation}{0}

\section{Introduction}

The celebrated AdS/CFT conjecture \cite{AdSCFT} connects string theory on an asymptotically $AdS_p \times S_q$ spacetime to a non-gravitational conformal field theory, in the original case $\mathcal{N} = 4$ superYang-Mills theory.   Over the last few years there has been a great deal of interest in extending this connection to field theories found in condensed matter systems.  Of particular interest has been the attempt, beginning with \cite{Kachru, Koroteev},  to study field theories with asymmetric scaling between spatial and time directions
\begin{equation} \label{scaling1}
t \rightarrow \lambda^z t, \hspace{5mm} \vec{x} \rightarrow \lambda \vec{x}
\end{equation}
known as Lifshitz field theories.  A more recent extension, known as hyperscaling violation, under the scaling (\ref{scaling1}) also introduces a constant conformal transformation on the spacetime metric
\be
ds^2 \rightarrow \lambda^{2 \theta/(d-2)} ds^2
\ee
for a $d$-dimensional bulk spacetime and a $d-1$-dimensional field theory, presuming as we will do throughout this work that $d \geq 3$.  From the field theory point of view a nonzero-$\theta$ involves a deviation of the scaling of entropy with respect to temperature from the naive dimensionful expectation \cite{Sachdev}
\be
S \sim T^{(d-2-\theta)/z}
\ee
The simplest guess for a gravitational dual is a spacetime of the form
\be \label{met0}
ds^2 = l^2 \Big(-r^{2+2 (d-2) (z-1)/(d-2-\theta)} dt^2+r^{-2-2 \theta/(d-2-\theta)} dr^2 + r^2 dx_i dx^i\Big)
\ee
which under the scaling above, as well as under
\be
r \rightarrow \lambda^{-1+\theta/(d-2)} r
\ee
reproduces the above scaling.  For the sake of compactness, however we will define $n_0$ and $n_1$ so that (\ref{met0}) becomes
\be \label{metdef}
ds^2 = l^2 \Big(-r^{n_0} dt^2 + \frac{dr^2}{r^{n_1}} + r^2 \delta_{i j} dy^i dy^j \Big)
\ee
where $\delta_{ij}$ is a flat metric.  It is worth remarking that while the above metrics with non-zero $\theta$ do not have any obvious interpretation asymptotically at large $r$--as is well known in the AdS/CFT literature a constant conformal transformation of the boundary metric is a non-normalizable deformation which alters the theory one is talking about--one might imagine spacetimes which have more traditional (e.g. AdS) asymptotics and have only such scalings in the interior of spacetimes.  In particular a variety of papers (see, for example \cite{Hyppapers}) have  discussed examples in string theory, most commonly near-brane limits, where one sees metrics of the above type with nonzero $\theta$.

In this paper we examine the singularity structure of this class of spacetimes.
After reviewing the conditions of remotely reasonable matter imposed by the null energy condition, we examine the conditions under which metrics of the above form (\ref{metdef}) possess traditional naked curvature singularities or perhaps less familiar null curvature singularities, which do not make any curvature invariant diverge and yet produce an infinite amount of tidal forces and pull apart any observer unfortunate enough to come close to them.   It turns out, besides pure AdS space, there is only a one parameter family of solutions which avoids either of the above types of severe singularities.  Unfortunately (as we shall discuss) this family requires a rather exotic matter content involving negative energy densities that cannot be interpreted as a cosmological constant and consequently, unless proven otherwise, would destabilize  the vacuum quantum mechanically, due to pair production of negative energy matter plus gravitons, radiation, or other conventional matter.   In particular, unless one is willing to admit divergent terms in one's Lagragian, the Einstein-scalar-Maxwell Lagrangian favored to date  in this subject \cite{Sachdev, Shaghoulian} cannot support this apparently regular family.

\setcounter{equation}{0}
\section{Energy conditions}

One may always consider metrics of any desired form to be a solution of general relativity, or any similar theory, provided one is willing to consider a stress-energy tensor of the form defined by the Einstein equations.   The principal benefit of the various energy conditions is to provide substantial restrictions on the types of metrics one might consider, since violating these conditions means one is forced to consider matter with some rather undesirable properties.   Perhaps the most intuitively obvious energy condition is the weak energy condition, the statement that given any timelike geodesic $\xi^a$
\be
G_{a b} \xi^a \xi^b = T_{a b} \xi^a \xi^b \geq 0
\ee
or in other words the statement that no observer can see a negative energy density.   Negative energy densities typically pose an extreme danger to one's theory, for unless such matter has no dynamical degrees of freedom (e.g. a cosmological constant) or a conservation law, boundary condition, or topological condition forbidding the excitation of these degrees of freedom, short of proof to the contrary, one can produce in an entirely unbounded fashion pairs of negative energy and positive energy (e.g. photons or gravitons) particles from the putative vacuum.   The closely related null energy condition, which in every sensible situation insofar as we are aware also implies  a violation of the weak energy condition, states that given any null vector $k^a$
\be \label{nullcond}
T_{\alpha \beta} k^{\alpha} k^{\beta} = G_{\alpha \beta} k^{\alpha} k^{\beta} = R_{\alpha \beta} k^{\alpha} k^{\beta} \geq 0
\ee
The null energy condition is generally believed to hold for any type of observed or reasonable matter, provided in dynamical situations one agrees to average over the geodesic to avoid the short time violations introduced by quantum mechanics (essentially the energy-time uncertainty relation).   In the present context we will only consider solutions (\ref{metdef})  satisfying the null energy condition and will be concerned if any violations of the weak energy condition cannot be described as due to a cosmological constant.

Given the desired metric (\ref{metdef})
\be
ds^2 = l^2 \Big(-r^{n_0} dt^2 + \frac{dr^2}{r^{n_1}} + r^2 \delta_{i j} dy^i dy^j \Big)
\ee
 via direct calculation one finds the nonzero components of the Ricci tensor are
\begin{eqnarray} \label{Ricciexpr}
R_{t t} &=& \frac{n_0}{4} r^{n_0+n_1-2} \Big[ n_0 + n_1 + 2 (d-3) \Big]  \nonumber \\
R_{r r} &=&  \frac{1}{4 r^2} \Big[ 2 n_0 - 2 (d-2) n_1 - n_0 (n_0 + n_1) \Big] \nonumber \\
R_{i j} &=& -\frac{\delta_{i j} r^{n_1}}{2} \Big[ n_0 + n_1 +2 (d-3) \Big]
\end{eqnarray}
For any null vector $k^{\alpha}$, 
\be \label{nullvec}
0 = k_{\alpha} k^{\alpha} = g_{t t} (k^t)^2 + g_{rr} (k^r)^2 + g_{i j} k^i k^j
\ee
or equivalently
\be
(k^t)^2 = r^{-n_0 - n_1} (k^r)^2 + r^{2 - n_0} \vec{k}^2
\ee
(where $\vec{k}^2 = k^i k_i$).   Hence the null energy condition in this context becomes the statement that
\begin{eqnarray}
0 \leq R_{\alpha \beta} k^{\alpha} k^{\beta} &=& 
 \frac{r^{n_1} \vec{k}^2}{4} (n_0 - 2) \Big[ n_0 + n_1 + 2 (d-3) \Big]  \nonumber \\
&+&
\frac{(d-2) (k^r)^2}{2 r^2} \Big[n_0 - n_1\Big]
\end{eqnarray}
and hence the null energy condition is equivalent to
\begin{eqnarray}
&I)& n_0 \geq n_1 \nonumber \\
&II)&  (n_0 - 2) \Big[ n_0 + n_1 + 2 (d-3) \Big]  \geq 0
\end{eqnarray}
The first condition ($n_0 \geq n_1$) will be quite useful for us in classifying the above metrics and the second will end up being automatic if one imposes the first as well as conditions that eliminate singularities.  For the record, we note the purpose of this section is mainly pedagogical and to clearly explain the importance of these conditions to those whose expertise lies in an area other than general relativity.  In particular, the restrictions implied by the null energy condition on metrics of the desired type have been previously given in \cite{Takayanagi, Kachruhyperscaling, Shaghoulian}.

\setcounter{equation}{0}
\section{Curvature Singularities}

The most familiar types of curvature singularities are those which involve the divergence of some scalar quantity calculated from the Riemann tensor (e.g. $R_{\alpha \beta \gamma \delta} R^{\alpha \beta \gamma \delta}$) which as a coordinate invariant guarantees the singularity is not merely due to a poor choice of coordinates but involves a pathology in the metric itself.   While it is relatively rare for such singularities to be absent from the square of the Riemann tensor and present in scalars constructed from more factors of the Riemann tensor, we would like a simple method for examining all possible such singularities universally. 

It turns out one can cary out this task rather simply by considering the components of the Riemann tensor in perhaps the simplest possible orthonormal basis.  Taking basis vectors essentially by taking the square roots of each of the metric components yields
\be \label{statmet}
(e_0)_{\alpha} = -l r^{{n_0}/2} \partial_{\alpha} t  \quad
(e_1)_{\alpha} =l r^{{-n_1}/2} \partial_{\alpha} r  \quad (e_i)_{\alpha} = l r  \partial_{\alpha} y_i  
\ee
Adopting the notation 
\be
R_{i j k l} \equiv R^{\alpha \kappa \gamma \delta} (e_i)_{\alpha}  (e_j)_{\kappa}  (e_k)_{\gamma}  (e_l)_{\delta}
\ee
the nonzero components of the Riemann tensor are
\begin{eqnarray}
R_{0101} &=& \frac{n_0 (n_0 + n_1 -2) r^{n_1-2}}{4 l^2}  \qquad 
R_{0i0j} = \frac{n_0 r^{n_1 - 2}}{2l^2} \delta_{i j}  \, \, \, \, (i \neq j) \nonumber \\
R_{1i1i} &=&- \frac{n_0 r^{n_1 - 2}}{2l^2} \delta_{i j}  \qquad 
R_{ijkl} = \frac{r^{n_1 - 2}}{l^2} (\delta_{i l} \delta_{j k} - \delta_{i k} \delta_{j l})  \nonumber
\end{eqnarray}
As a result, curvature invariants will diverge as $r \rightarrow 0$ (i.e. the deep interior) unless $n_1 \geq 2$.   Conversely, at large $r$ curvature invariants will diverge unless $n_1 \leq 2$.   In terms of a holographic interpretation, such large curvatures asymptotically would be disastrous (in no sense would gravity become decoupled) and even those advocating the study of these metrics frequently imagine such metrics flowing to AdS asymptotics.  A curvature divergence in the deep interior is not necessarily disastrous--some may be cured by $\alpha'$ effects in string theory or other effects from quantum gravity, but, as we will discuss below, provided one does not violate the null energy condition signals can reach observers at finite $r$  from $r=0$ in finite affine parameter. Hence a divergence of this type at $r= 0$ is a naked singularity where Planck-scale physics will affect distant observers and classical calculations are not generically reliable.   The only metrics of the type (\ref{metdef}) avoiding the divergence of curvature invariants everywhere then are those with $n_1 = 2$, in which case one returns to the previously studied Lifshitz case.

 Now considering the geodesics of the relevant spacetime, the metric (\ref{metdef}) has a timelike killing vector and $d-2$ spacelike killing vectors, resulting in the conserved energies and momenta 
\be
E = -g_{t t} \dot{t}
\ee
\be
p_i = g_{i i} \dot{y}_i
\ee
with
\be
\dot{x}^{\mu} = \frac{d x^{\mu}}{d\lambda}
\ee
for some affine parameter $\lambda$. Then for a geodesic
\be
-k_0 = g_{t t} \dot{t}^2+g_{r r} \dot{r}^2+\Sigma_i g_{y_i y_i} \dot{y}_i^2 
\ee
that is either timelike ($k _0= 1$) or null ($k_0 = 0$)  we have 
\be \label{geod1}
\dot{r}^2 = \frac{r^{n_1}}{l^2} \Bigg( \frac{E^2}{l^2} r^{-n_0} -\frac{\vec{p}^2}{l^2 r^2} - k_0\Bigg)
\ee
Note in particular that for null geodesics without any transverse momentum
\be
\dot{r}^2 r^{n_0-n_1} = \frac{E^2}{l^4}
\ee
and so as long as $n_0$ is not overly negative, and in particular if the null energy condition is obeyed ($n_0 \geq n_1$), then
\be
r^{\frac{n_0}{2}-\frac{n_1}{2}+1} = \pm \frac{E}{2 l^2} (n_0-n_1+2) (\lambda - \lambda_0)
\ee
and geodesics travel from $r = 0$ to finite $r$ in finite affine parameter $\lambda$.

For $n_0 \leq 0$ the surface $r = 0$ is a timelike surface. As long as one enforces the null energy condition, $0 \geq n_0 \geq n_1$, yielding a familiar naked singularity with the same Penrose diagram as a negative mass black hole, as illustrated in Figure 1. For $n_0 > 0$, the surface $r = 0$ is a null surface and so the Penrose diagram for this metric looks like the Poincare patch in AdS  as in Figure 2, possibly with a singularity along the $r =0$ surface.  Note that one can receive signals from the past surface at $r = 0$ (what would be a Poincare horizon in the AdS case) even in the  Poincare-type patch and, just like AdS, any timelike observer will eventually run into $r = 0$ (\ref{geod1}).  Hence it appears any singularities of this type should be regarded as dangerous.  This contrasts with the rather more familiar situation involving null curvature singularities in asymptotically flat space where, provided one can create the singularity from some smooth initial data via collapse or other physical process and hence eliminate the past singularity, signals from the null singularity can not be received at any finite time, affording the option of  regarding the singularity as not particularly problematic.
\begin{figure}
\centering
    
\includegraphics[scale= 0.6]{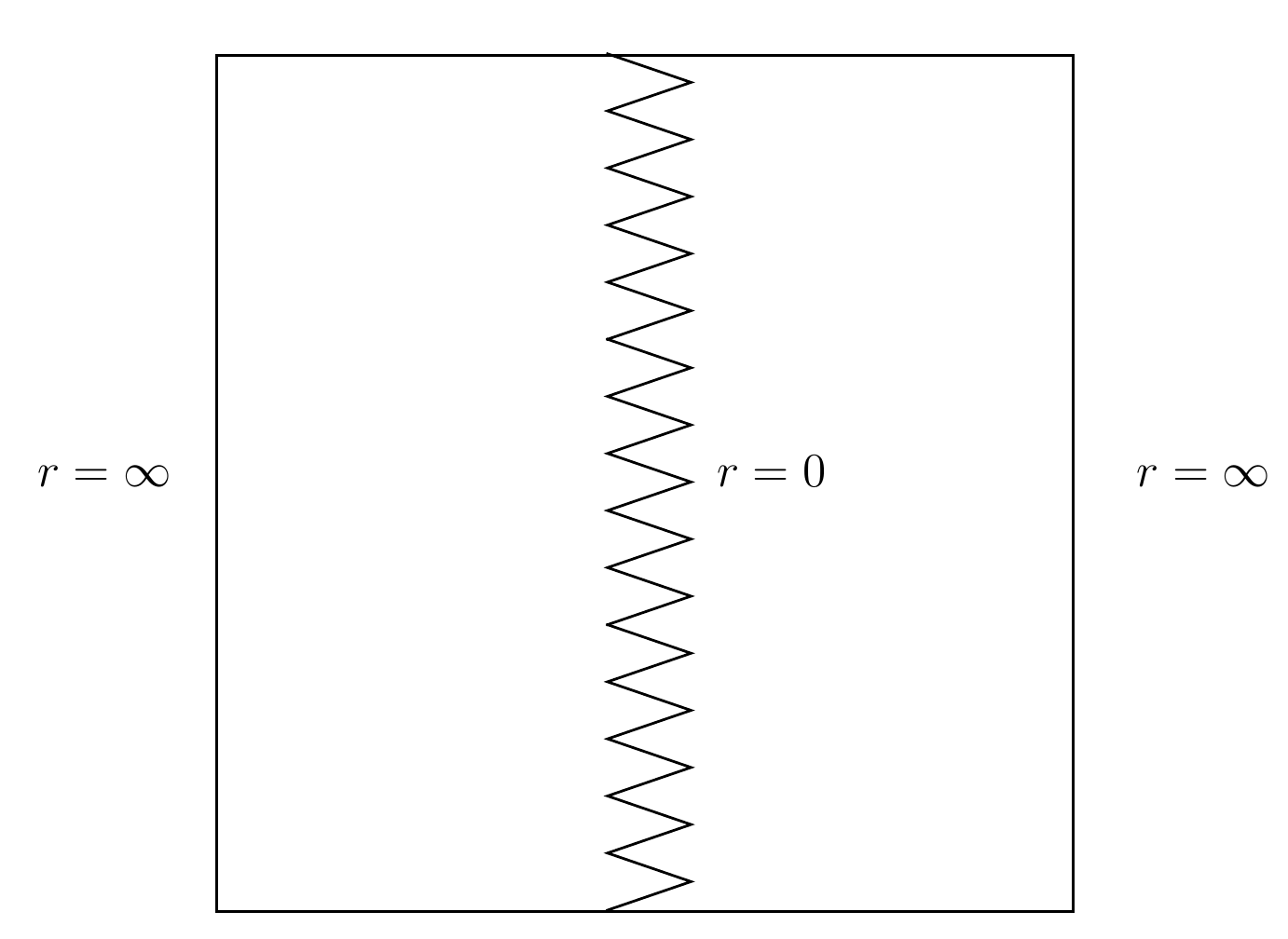}
	\caption{Penrose diagram for $n_0 < 0$}
	\label{ex1fig}
	\end{figure}
	
	\begin{figure}
	\centering
    
\includegraphics[scale= 0.6]{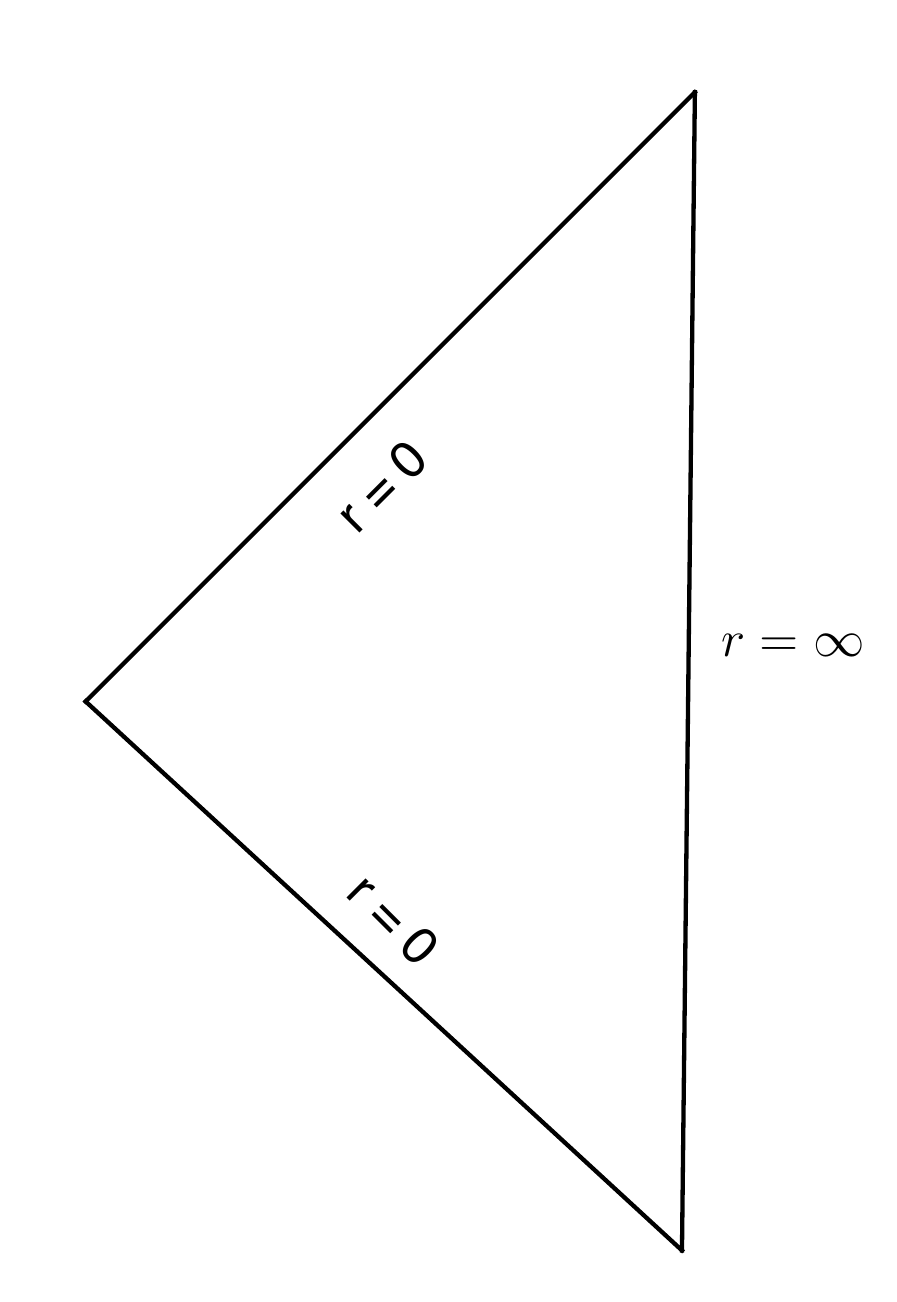}
	\caption{Penrose diagram for $n_0 \geq 0$}
	\label{ex1fig2}
	\end{figure}

One of the remarkable properties of null curvature singularities is that they do not necessarily make any curvature invariant diverge.  The null curvature singularity in the original Lifshitz spacetime \cite{Kachru}, as well as in singular gravitational plane waves \cite{HorowitzSteif} and a variety of other examples \cite{HorowitzRossNaked}, is of this type.  Historically such singularities were frequently described within the general relativity literature as ``mild'', in comparison to other types of singularities where curvature invariants diverge.   From the point of view of string theory, however, this classification is essentially entirely backwards.   It appears $\alpha'$ corrections must cure certain singularities involving the divergence of curvature invariants, (e.g. those in 0, 1, and 2-branes) and at any rate near such singularities the classical theory is insufficient and $\alpha'$ corrections are important.  On the other hand, given a singularity where all curvature invariants remain small, $\alpha'$ corrections remain negligable\footnote{One might worry that matter fields coupled to the metric could become large near such a singularity and introduce significant $\alpha'$ corrections.  To exploit this loophole, however, such matter fields would somehow have the rather odd property of becoming large without producing a large stress tensor (which of course would be reflected in the metric). We know of no such cases and presumably such a scenario from a quantum mechanical point of view would involve many quanta at low energies and hence be problematic for the reasons familiar from the consideration of black hole remnants.} and if, in addition, the dilaton does not become large near the singularity, the supergravity approximation remains a good one and the solution apparently should be regarded as singular in string theory as well as classically.   

The way to test for the presence of curvature singularities not present in a curvature invariant is examine the components of the Riemann tensor in a parallely-propagated-orthonormal-frame (PPON)--that is, the Riemann tensor as measured by an observer freely falling along a geodesic.
Constructing a PPON with a unit timelike vector parallel to the four velocity along a timelike geodesic with conserved energy $E$ and transverse momentum $p$ which, without loss of generality, we may take to be pointing along a particular $y$ direction, say $y_1$, we have:
\begin{eqnarray}
(\tilde{e}_0)_{\alpha} &=& -E  \partial_{\alpha} t \pm l r^{-(n_0+n_1)/2} \sqrt{\frac{E^2}{l^2}- r^{n_0} \Big(1 + \frac{p^2}{l^2 r^2} \Big)}  \,  \partial_{\alpha} r + p \, \partial_{\alpha} y_1 \nonumber \\
(\tilde{e}_1)_{\alpha} &=& \beta_1 \, \partial_{\alpha} t + \beta_2 \, \partial_{\alpha} r  \\
(\tilde{e}_2)_{\alpha} &=& \gamma_1 \, \partial_{\alpha} t + \gamma_2 \, \partial_{\alpha} r  + \gamma_3 \, \partial_{\alpha} y_1  \nonumber \\
(\tilde{e}_i)_{\alpha} &=& l r  \partial_{\alpha} y_i   \nonumber 
\end{eqnarray}
where the two choices of sign correspond to whether one is considering a radially ingoing or outgoing geodesic and $i \geq 2$.  The constants $(\beta_1, \beta_2, \gamma_1, \gamma_2, \gamma_3)$ are uniquely determined by requiring the $\tilde{e}_{\alpha}$ to be an orthonormal basis (provided one, as we do, chooses one basis vector to have no component along the $y_i$); we omit their explicit form as they are slightly messy and unilluminating. 
Then, using a notation analogous to the above,
\be
\tilde{R}_{i j k l} \equiv R^{\alpha \kappa \gamma \delta} (\tilde{e}_i)_{\alpha}  (\tilde{e}_j)_{\kappa}  (\tilde{e}_k)_{\gamma}  (\tilde{e}_l)_{\delta}
\ee
(i.e. the components in a PPON frame) we obtain 
\begin{eqnarray}
\tilde{R}_{0101} &=& \frac{r^{n_1-n_0 -2}}{4 l^4 (p^2+l^2 r^2)} \Big[ 2 (n_0 - n_1) E^2 p^2 \nonumber \\
&+& n_0 r^{n_0-2} (p^2 + l^2 r^2) [ (n_1 + n_0 - 4) p^2 + (n_1 + n_0 - 2) l^2 r^2] \Big] \nonumber \\
\tilde{R}_{0102} &=& -\frac{(n_0 - n_1)E p \, r^{n_1-n_0-1} \sqrt{\frac{E^2}{l^2}- r^{n_0} \Big(1+\frac{p^2}{l^2 r^2} \Big)}}{2 l^2 (p^2 + l^2 r^2)} \nonumber \\
\tilde{R}_{0112} &=& \frac{p \, r^{n_1-n_0-2} [2 (n_1  - n_0) E^2 - n_0 (n_0+n_1-4) r^{n_0 - 2} (p^2 + l^2 r^2)]}{4 l^4 \sqrt{p^2 + l^2 r^2}} \nonumber \\
\tilde{R}_{0202} &=& \frac{r^{n_1 - n_0} [ (n_0 - n_1) E^2 +n_1 r^{n_0 - 2} (p^2+l^2 r^2)]}{2 l^2 (p^2 + l^2 r^2)} \nonumber \\
\tilde{R}_{0212}&=& \frac{(n_0 - n_1)E  \, r^{n_1-n_0-1} \sqrt{\frac{E^2}{l^2}- r^{n_0} \Big(1+\frac{p^2}{l^2 r^2} \Big)}}{2 l^2 (p^2 + l^2 r^2)} \nonumber \\
\tilde{R}_{0i0j} &=& \delta_{i j} \frac{r^{n_1-n_0 - 2} \Big[ (n_0 - n_1) E^2 + r^{n_0-2} [ (n_1-2) p^2 + n_1 l^2 r^2) ]\Big]}{2 l^4} \nonumber \\
\tilde{R}_{0i1j} &=& \delta_{i j} \frac{(n_0 - n_1) E r^{n_1-n_0-1}\sqrt{\frac{E^2}{l^2}- r^{n_0} \Big(1+\frac{p^2}{l^2 r^2} \Big)}}{2 l^6 (p^2 + l^2 r^2)} \nonumber \\
\tilde{R}_{0i2j} &=& \delta_{i j} \frac{p \, r^{n_1-n_0-2} [(n_0 - n_1) E^2 + (n_1 - 2) r^{n_0-2} (p^2 + l^2 r^2)]}{2 l^4 \sqrt{p^2 + l^2 r^2}} \nonumber \\
\tilde{R}_{1212} &=& \frac{r^{n_1 -n_0-2}\Big[ 2 (n_0 - n_1) E^2 + n_0 r^{n_0-2} \Big( (n_1 + n_0- 4) p^2 - 2 l^2 r^2) \Big)\Big]}{4 l^4} \nonumber \\  
\tilde{R}_{1i1j} &=& \delta_{i j} \frac{r^{n_1-n_0} \Big[ (n_0 - n_1) E^2-n_0 r^{n_0-2} (p^2 + l^2 r^2) \Big]}{2 l^2 (p^2 + l^2 r^2)} \nonumber \\
\end{eqnarray}
\begin{eqnarray}
\tilde{R}_{1i2j} &=& \delta_{i j}  \frac{(n_0 - n_1) p E \, r^{n_1 - n_0-1} \sqrt{\frac{E^2}{l^2}- r^{n_0} \Big(1+\frac{p^2}{l^2 r^2} \Big)}}{2 l^2 (p^2 + l^2 r^2)} \nonumber \\
\tilde{R}_{2i2j} &=& \delta_{i j} \frac{r^{n_1-n_0 - 2} \Big[(n_0 - n_1) p^2 E^2 + r^{n_0-2} \Big( (n_1-2) p^2 - 2 l^2 r^2 \Big) \Big(p^2 + l^2 r^2 \Big) \Big]}{2 l^4 (p^2 + l^2 r^2)} \nonumber \\
\tilde{R}_{ijkl} &=& \frac{r^{n_1 - 2}}{l^2} (\delta_{i l} \delta_{j k} - \delta_{i k} \delta_{j l})\end{eqnarray}
Provided we insist that as $r \rightarrow 0$ no curvature invariants diverge $(n_1 \geq 2)$ and that the null energy condition is not violated $(n_0 \geq n_1)$, we find
\be
n_1 - n_0 \leq 0
\ee
and $n_0 \geq n_1 \geq 2$.   Then various components of the Riemann tensor in an orthonormal basis (e.g. $\tilde{R}_{0202}$) will diverge as $r \rightarrow 0$ even if $p = 0$ unless
\be
n_0 = n_1
\ee
All the components of the Riemann tensor will be finite as $r \rightarrow 0$ if and only if either
\be
n_0 = n_1 = 2
\ee
that is, pure AdS, or
\be \label{regsol}
n_1 = n_0 \geq 4
\ee
The remaining part of the null energy condition given $n_0 = n_1$ is the criterion that
\be
(n_0 - 2) (n_0 + d - 3) \geq 0
\ee
which is clearly satisfied in either of the above cases.  We note for the record \cite{Shaghoulian} examined tidal forces in these spacetimes in the absence of transverse momentum $p$ and noted the existence of an apparently regular one-parameter family solutions, a subset of which fall into the class of solutions (\ref{regsol}) without obvious singularities if one considers geodesics with general $p$.
\setcounter{equation}{0}
\section{Solution Classification}

As a result of the above analysis, provided one does not violate the null energy condition one is left with four classes of solutions
\begin{eqnarray}
&I)& n_1 < 2 \, \, (R_{\alpha \beta \gamma \delta} R^{\alpha \beta \gamma \delta} \, \, \mathrm{divergent} \, \, \mathrm{at} \, r = 0) \nonumber \\
&II)& n_0 = n_1 = 2 \, \, \, (\, \mathrm{pure}\, \,  \mathrm{AdS}) \nonumber \\
&III)& n_0 = n_1 \geq 4 \nonumber \\
&IV)& n_0 > n_1 \geq 2 \, \, \mathrm{or} \, \, 2 < n_0= n_1 < 4 \, \, (\mathrm{null} \, \,  \mathrm{curvature} \, \, \mathrm{singularity}\, \, \mathrm{at} \, r = 0) \nonumber
\end{eqnarray}
that is, besides the pure AdS solution ($II$), one has a familiar naked singularity ($I$) or a perhaps less familiar null curvature singularity ($IV)$, leaving only a one parameter family of solutions ($III$) that is not obviously problematic. 

To compare these categories with previous results, it is useful to provide a translation into the parameters used in various different sources--regrettably several different conventions have emerged.  In the notation of Huijse, Sachdev and Swingle \cite{Sachdev} where the metric is written
\be
ds^2 = \frac{1}{u^2} \Big( -\frac{dt^2}{u^{2 d_t (z - 1)/(d_t -\theta)}} + u^{2 \theta/(d_t - \theta)} du^2 + dx_i^2 \Big)
\ee
and $d_t = d-2$ is the number of transverse spatial directions $x_i$, we have
\be
z = \frac{n_0 + n_1 - 2}{n_1},  \, \, \, \, \, \, \, \, \theta = ( d - 2) \Big(1 - \frac{2}{n_1} \Big)
\ee
and using those results in the above notation the area scaling law for entanglement entropy will be satisfied if
\be \label{entscaling}
2 (d - 2) \geq n_1 > 0
\ee
In higher dimensions (\ref{entscaling}) becomes less restrictive but in four dimensions the only case (aside from pure AdS) that avoids singularities at $r = 0$ is $n_0 = n_1 = 4$.  Notably, this is precisely the case where good agreement with the field theory results was obtained \cite{Sachdev}.

On the other hand, the  type ($III$) solutions turn out to require a somewhat exotic stress-energy tensor.   For any metrics of the type (\ref{metdef}) the corresponding energy density in the static frame (\ref{statmet}) is 
\be \label{rhostat}
\rho = G_{\alpha \beta} (e_0)^{\alpha} (e_0)^{\beta} = -\frac{(d-2) (d- 3 + n_1)}{2 l^2} r^{n_1-2}
\ee
and -- unless $n_1 \leq - (d-3)$ (and one has the naked timelike singularities in class ($I$)) --  $\rho < 0$,   violating  the weak energy condition.  As discussed before, any matter source that has dynamical degrees of freedom violating the weak energy condition is generically dangerous at least from a quantum mechanical point of view--in the absence of a symmetry, conserved charge, or other argument to forbid it one expects a rapid pair production of negative energy quanta and positive energy quanta (e.g. radiation) from the putative vacuum.   On the other hand a source which has no dynamics of its own, most notably a negative cosmological constant, does not present fundamental problems.   In the case of $n_1 = 2$, $\rho$ is a constant and this violation may be attributed to a cosmological constant (as, for example, in the pure AdS case) and one need not be disturbed about this violation.   More broadly, any stress tensor which is a conventional (positive energy density) term plus a negative cosmological constant need not disturb us in this regard.  On the other hand, if $n_1 \neq 2$ (and $n_1> -(d-3)$) the negative energy density is necessarily dynamical, increasing in magnitude with increasing $r$ if $n_1 > 2$ and diverging as $r \rightarrow 0$ if $2 > n_1 > - (d-3)$.   This would appear to indicate one must regard many of the type ($I$) solutions as deeply pathological--i.e. negative mass singularities that had better be excluded from any sensible theory- and one only has left in this class the rather exotic naked timelike singularities where $n_1 \leq -(d-3)$.   In the case $n_1 > 2$ the negative energy density grows at increasing $r$ and one may well object that at sufficiently large $r$ the metric surely  cannot be trusted.   Perhaps more fundamentally, however, it is difficult to understand how one can support any solutions of this type, including those of class ($III$), without running into the exotic matter problems described above.

For completeness, we also list the pressures as
\be
p_r =  G_{\alpha \beta} (e_1)^{\alpha} (e_1)^{\beta} =  \frac{(d-2) (d- 3 + n_0)}{2 l^2} r^{n_1-2}
\ee
and 
\be
p_x = G_{\alpha \beta} (e_2)^{\alpha} (e_2)^{\beta} =\Big[ n_0 (n_0+n_1) +2 (d-4) n_0 + 2 (d-3) n_1 + 2 d^2-14 d + 24 \Big] \frac{r^{n_1-2}}{4 l^2}
\ee
It is a rather odd fact that if $n_1 > 2$ all these energy densities and pressures go to zero as one approaches $r = 0$--that is that if there is a smooth continuation through $r = 0$ apparently the solution approaches a vacuum.  On the other hand, it is easy to check that the only Ricci flat solution of the desired form is the nakedly singular $n_0 = n_1 = 3-d$.

\setcounter{equation}{0}
\section{Phenomenological Lagrangians}

The above suggests there well may not be sensible stress energy tensors giving rise to the case ($III$) metrics.   However, let us examine in detail the commonly explored phenomenological Lagrangians giving rise to these forms of metrics and in particular the  Lagrangian
\be
S = \kappa \int \sqrt{-g} \Big(R - \frac{1}{2} (\nabla \phi)^2 - \frac{\alpha(\phi)}{4} F_{a b} F^{a b} - V(\phi) \Big)
\ee
yielding
\be
R_{a b} = \frac{1}{2} \nabla_a \phi \nabla_b \phi + \frac{\alpha(\phi)}{2} F_{a c} F_{b}^{c} +\frac{g_{a b}}{d-2} \Big( V(\phi) - \frac{\alpha(\phi)}{4} F_{c d} F^{c d} \Big)
\ee
Given the symmetries of the desired metrics, without loss of generality we may take the only nonzero component of the field to be $F_{t r}(r)$ and, in the case of a four dimensional solution, a magnetic field $F_{x_1 x_2} = Q_1$ for some constant $Q_1$.  We may then algebraically solve the Ricci equations (\ref{Ricciexpr}) with the result
\begin{eqnarray} \label{Riccisol}
{\phi'(r)}^2 &=& \frac{(d-2) (n_0 - n_1)}{r^2} \nonumber \\
\alpha(\phi) (F_{t r})^2 &=& \frac{l^2}{2} \Big(n_0 - 2 \Big)  \Big( n_0 + n_1 + 2 (d-3) \Big) r^{n_0- 2} -Q_1^2 \alpha(\phi) r^{n_0-n_1-4} \nonumber \\
V(\phi) &=& - \frac{(n_0+ n_1 +2 ( d - 3) ) (n_0 + 2 (d-3)) r^{n_1 - 2}}{4 l^2}
\end{eqnarray}
In the case $n_0 = n_1$, then $\phi'(r)^2 = 0$ and $\phi$ is necessarily a constant.   Further,  in the absence of poles or other singularities at finite $\phi$, this implies that $\alpha(\phi)$ is simply a constant in any region where $n_0=n_1$.   In this case the equation for the field
\be
\nabla_a \Big( \alpha(\phi) F^{a b} \Big) = 0
\ee
becomes
\be
\partial_r \Big(\sqrt{-g} F^{r t} \Big) = 0
\ee
and hence
\be \label{fieldeqn}
F_{t r} = Q_0 r^{2-d}
\ee
for some constant $Q_0$.  Matching powers of $r$   on both sides of the second equation in (\ref{Riccisol}), we find that (\ref{fieldeqn}) implies
\be
n_0 = 6 - 2 d
\ee
(In the case of nonzero $Q_1$, the positivity of the left hand side of the second equation in (\ref{Riccisol}) ensures the magnetic term is never dominant  as $r \rightarrow 0 $ and hence $n_0 - 2 \leq n_0- n_1 - 4 = -4$.)   Consequently $n_1 = n_0 \leq 0$ and one has a a class ($I$) singular solution.  Alternatively, one might suppose that both sides of the second equation in (\ref{Riccisol}) vanish identically and hence not only $Q_0 = 0$ but  
\be
Q_1^2 \alpha(\phi) = \frac{l^2}{2} (n_0 - 2) (n_0 + n_1 + 2 (d-3)) r^{n_1+2}
\ee
which in turn implies that aside from the naked singularity case ($I$) of
\be
n_0 = n_1 = -2
\ee
either
\be
n_0 = 2
\ee
i.e. $n_1 = n_0 = 2$ and $Q_0 = Q_1 = 0$, (that is, a pure AdS solution) or
\be
0 = (n_0 + d-3) = Q_1
\ee
and one has the Ricci-flat naked singularity mentioned before $(n_0 = 3-d)$.  Thus this Lagragian simply does not allow any type ($III$) solutions provided there are no poles or singularities at finite $\phi$ in $\alpha(\phi)$.  

On the other hand, if one allowed $\alpha(\phi)$ to diverge at some finite $\phi_c$, then one could arrange for $\phi \rightarrow \phi_c$ as $r \rightarrow 0$ and evade the above constraints.   Indeed, if one generalizes the above procedure to a more general metric that only approaches a class $(III)$ solution as $r \rightarrow 0$ and algebraically solves the Einstein equations for $(\phi'(r)^2$, $\alpha(\phi)$, and $V(\phi)$ as above then $\alpha(\phi)$ will necessarily diverge.  In particular, it is straightforward to check that the ``regular'' solution of \cite{Shaghoulian} which followed such a procedure is of this type and in particular requires a sixth order pole at finite $\phi$ in $\alpha(\phi)$.   We take the perspective that any ``solution'' relying  on a truly divergent Lagrangian can not be considered sensible.

The other phenomenological Lagrangian familiar from the study of Lifshitz ($n_1 = 2$) solutions is a massive vector field together with a cosmological constant
\be
S = \kappa \int \sqrt{-g} \Big(R- \frac{1}{4} F_{a b} F^{a b} -\frac{m_0^2}{2} A_a A^a - 2 \Lambda \Big)
\ee
yielding
\be
R_{a b} = \frac{1}{2} F_{a c} F_{b}^c + \frac{m_0^2}{2} A_a A_b + \frac{g_{a b}}{d-2} \Big( 2 \Lambda - \frac{F_{c d} F^{c d}}{4} \Big)
\ee
As usual in such studies taking only a purely electric vector field\footnote{Any non-zero magnetic field necessitates a dependence of the potential on $y_i$ and it appears inevitable that for nonzero $m_0$ this dependence will break the killing symmetries assumed in the metric (\ref{metdef}).}   $A_t(r)$ again solving the Ricci equations algebraically one finds 
\be
2 \Lambda = k_0 r^{n_1 - 2}
\ee
for a slightly messy constant $k_0$.   Then one must either take $n_1 = 2$, in which case one returns to the Lifshitz case and the null curvature singularities, as well as other difficulties, involved there \cite{CopseyMann} or a zero cosmological constant.  If we seek solutions of the type $n_0 = n_1$ one finds
\begin{eqnarray}
2 \Lambda &=& -\frac{ \Big( n_0 + d - 3\Big) \Big(n_0 + 2 (d-3) \Big) r^{n_0 - 2}}{2 l^2} \nonumber \\
(F_{t r})^2 &=& l^2 (n_0 - 2) (n_0 + d - 3) r^{n_0 - 2}
\end{eqnarray}
and besides the pure AdS solutions $(n_0 = n_1 = 2)$ one is forced to take $\Lambda = 0$ and hence either
\be
n_0 = 3 - d
\ee
in which case the field vanishes and one again finds the class ($I$) Ricci-flat singular solution mentioned above or
\be
n_0 = 2 (3 -d)
\ee
in which case one finds the remaining Einstein equations yield
\be
(F_{t r})^2 = 2 l^2 (d-2) (d-3) r^{4 - 2 d}
\ee
and 
\be
m_0^2 A_t^2= 0
\ee
and,  besides the singular class ($I$) vacuum solutions for $d=3$, hence one has a solution only if the vector field is the usual massless Maxwell-field after all. Even in that case one has a naked timelike singularity of the type $(I)$, albeit a charged one in this case.

We of course are not in a position to consider all possible Lagrangians.   As discussed above, due to the exotic nature of the matter required to support the only regular class of solutions one should expect generically if one tries to force conventional matter to serve in this role either a divergence in the field or in the Lagrangian probably should have been expected from the beginning.   In particular, any Lagrangian like the Einstein-Maxwell-scalar one with a conserved vector field is particularly problematic simply due to Gauss's law--one might try to construct a ``ground state'' out of an extremal black hole but if one does not allow any source of a non-trivial field a singularity at some point is essentially inevitable.   Such considerations do not, of course, rule out the possibility one might find some configuration, quite likely some kind of condensate, with an energy density which goes negative but which is bounded from below by considerations of topology or boundary conditions.

\section*{Acknowledgements}
It is a pleasure to thank  S. Sachdev for useful discussions.  This work was supported in part by the Natural Sciences and Engineering Research Council of Canada.   Research at Perimeter Institute is supported by the Government of Canada through Industry Canada and by the Province of Ontario through the Ministry of Research \& Innovation.

\end{document}